\documentclass[preprint,prd,superscriptaddress,showpacs,floatfix,preprintnumbers,nofootinbib]{revtex4-1}
\usepackage{color}
\usepackage{graphics}
\usepackage{amsmath, amssymb, epsfig}
\usepackage{subfigure}
\newcommand{\mathsym}[1]{{}}

\def\lsim{\:\raisebox{-1.1ex}{$\stackrel{\textstyle<}{\sim}$}\:}
\def\gsim{\:\raisebox{-1.1ex}{$\stackrel{\textstyle>}{\sim}$}\:}

\newcommand{\ba}{\begin{array}}
\newcommand{\ea}{\end{array}}
\newcommand{\bal}{\begin{align}}
\newcommand{\eal}{\end{align}}
\newcommand{\be}{\begin{equation}}
\newcommand{\ee}{\end{equation}}
\newcommand{\beqa}{\begin{eqnarray}}
\newcommand{\eeqa}{\end{eqnarray}}
\newcommand{ \mysmall}[1]{\scriptscriptstyle #1} 
\newcommand{ \eq}[1]{Eq.~(\ref{#1})}
\def\321{$SU(3)\times SU(2)\times U(1)$}

\def\ca{\cos\alpha}
\def\cb{\cos\beta}
\def\sa{\sin\alpha}
\def\sb{\sin\beta}

\begin{document}

\title{Limiting two-Higgs-doublet models}

\author{Alessandro Broggio}
\email{alessandro.broggio@psi.ch}
\affiliation{Paul Scherrer Institut, CH-5232 Villigen, Switzerland}

\author{Eung Jin Chun}
\email{ejchun@kias.re.kr }
\affiliation{Korea Institute for Advanced Study, Seoul 130-722, Korea}

\author{Massimo Passera}
\email{passera@pd.infn.it}
\affiliation{Istituto Nazionale Fisica Nucleare, Sezione di Padova, I-35131 Padova, Italy}

\author{Ketan M.~Patel}
\email{ketan.patel@pd.infn.it}
\affiliation{Istituto Nazionale Fisica Nucleare, Sezione di Padova, I-35131 Padova, Italy}

\author{Sudhir K.~Vempati}
\email{vempati@cts.iisc.ernet.in}
\affiliation{Centre for High Energy Physics, Indian Institute of Science, Bangalore 560012, India}

\begin{abstract}
We update the constraints on two-Higgs-doublet models (2HDMs) focusing on the parameter space relevant to explain the present muon $g$$-$$2$ anomaly, $\Delta a_{\mu}$, in four different types of models, type I, II, ``lepton specific" (or X) and ``flipped" (or Y).  
We show that the strong constraints provided by the electroweak precision data on the mass of the pseudoscalar Higgs, whose contribution may account for $\Delta a_{\mu}$, are evaded in regions where the charged scalar is degenerate with the heavy neutral one and the mixing angles $\alpha$ and $\beta$ satisfy the Standard Model limit $\beta-\alpha \approx \pi /2$.
We combine theoretical constraints from vacuum stability and perturbativity with direct and indirect bounds arising from collider and $B$~physics. Possible future constraints from the electron $g$$-$$2$ are also considered.
If the 126 GeV resonance discovered at the LHC is interpreted as the light CP-even Higgs boson of the 2HDM, we find that only models of type X can satisfy all the considered theoretical and experimental constraints. 
\end{abstract}
\preprint{PSI-PR-14-05}
\preprint{KIAS-P14047} 

 
\maketitle

\section{Introduction}

The ATLAS and CMS Collaborations at the LHC~\cite{Aad:2012tfa,Chatrchyan:2012tx} found a neutral boson with a mass of about 126 GeV which confirms  the Brout-Eglert-Higgs mechanism. It is now of imminent interest to check whether this new boson is the unique one following exactly the Standard Model (SM) prediction, or if there are other bosons participating in the electroweak (EW) symmetry breaking.  One of the simplest way to extend the SM is to consider two Higgs doublets participating in the EW symmetry breaking instead of the standard single one. There are in fact several theoretical and experimental reasons to go beyond the SM and look forward to non-standard signals at the next run of the LHC and at future collider experiments. For reviews on two-Higgs-doublet models, see \cite{Gunion:1989we,Branco:2011iw}.

A major constraint to construct  models with two Higgs doublets (2HDMs) arises from flavour changing neutral currents, which are typically ubiquitous in these models. Requiring Natural Flavour Conservation (NFC) restricts the models to four different classes which differ by the manner in which the Higgs doublets couple to fermions~\cite{Gunion:2002zf,Aoki:2009ha,Branco:2011iw}. They are organized via discrete symmetries like $Z_2$ under which different matter sectors, such as right-handed leptons or left-handed quarks, have different charge assignments. These models  are labeled as type I, II, ``lepton-specific" (or X) and ``flipped" (or Y). Normalizing the Yukawa couplings of the neutral bosons in such a way that the explicit Yukawa interaction terms in the Lagrangian are given by $(y^{\phi}_f) \frac{m_f}{\upsilon} \bar{f} f \phi$ for the CP-even scalars $\phi=h, H$ (lighter and heavier, respectively) and $i (y^A_f) \frac{m_f}{\upsilon}\bar{f} \gamma_5 f A$ for the pseudoscalar $A$ in the mass eigenstate basis, the $y^{h,H,A}_f$ factors are summarized in Table I for each of these four types of 2HDMs as functions of $\tan\beta\equiv v_2/v_1$, the ratio of the two Higgs vacuum expectation values, and the diagonalization angle $\alpha$ of the two CP-even Higgs bosons ($v=\sqrt{v_1^2 + v_2^2}= 246$~GeV). It should be noted however, that  in addition to these models, NFC can also occur in models with alignment, as in~Ref.~\cite{Pich:2009sp}. In this class of models, more general sets of relations are imposed on the field content using discrete symmetries similar to $Z_2$, which still conserve flavour but allow for CP violation. A class of 2HDMs also exists where one of the Higgs doublets does not participate in the dynamics and remains \textit{inert}~\cite{Ma:2006km,Barbieri:2006dq}. Finally, in the so-called type III models both up and down fermions couple to both Higgs doublets. A detailed analysis of flavour and CP violation in type III models can be found in~\cite{Crivellin:2013wna} and references therein.  
%
\begin{table}[!ht]
\begin{small}
\begin{center}
\begin{tabular}{|l|ccccccccc|}
 \hline
 \hline
~~~~~~~~&~~~$y_u^A$~~~ & ~~~$y_d^A$~~~ & ~~~$y_l^A$~~~ & ~~~~$y_u^H$~~~ & ~~~$y_d^H$~~~ & ~~~$y_l^H$~~~ & ~~~$y_u^h$~~~ & ~~~$y_d^h$~~~ & ~~~$y_l^h$~~~\\
\hline
~Type I~~ &$\cot\beta$ & $-\cot\beta$ & $-\cot\beta$ & $\frac{\sa}{\sb}$ & $\frac{\sa}{\sb}$ & $\frac{\sa}{\sb}$ & $\frac{\ca}{\sb}$ & $\frac{\ca}{\sb}$ & $\frac{\ca}{\sb}$~ \\
~Type II~~ &$\cot\beta$ & $\tan\beta$ & $\tan\beta$ & $\frac{\sa}{\sb}$ & $\frac{\ca}{\cb}$ & $\frac{\ca}{\cb}$ & $\frac{\ca}{\sb}$ & $-\frac{\sa}{\cb}$ & $-\frac{\sa}{\cb}$~ \\
~Type X~~ &$\cot\beta$ & $-\cot\beta$ & $\tan\beta$ & $\frac{\sa}{\sb}$ & $\frac{\sa}{\sb}$ & $\frac{\ca}{\cb}$ & $\frac{\ca}{\sb}$ & $\frac{\ca}{\sb}$ & $-\frac{\sa}{\cb}$~ \\
~Type Y~~ &$\cot\beta$ & $\tan\beta$ & $-\cot\beta$ & $\frac{\sa}{\sb}$ & $\frac{\ca}{\cb}$ & $\frac{\sa}{\sb}$ & $\frac{\ca}{\sb}$ & $-\frac{\sa}{\cb}$ & $\frac{\ca}{\sb}$~ \\
\hline
\hline
\end{tabular}
\end{center}
\end{small}
\caption{The normalized Yukawa couplings of the neutral bosons to up- and down-type quarks and charged leptons.}
\label{yukawas}
\end{table}

One of the possible experimental indications for new physics is the measurement of the muon $g$$-$$2$ ($a_{\mu}$), which at present shows a 3--3.5$\sigma$ discrepancy $\Delta a_{\mu}$ from the SM prediction. Although not large enough to claim new physics, it can be used as a guideline to single out favourable extensions of the SM.  In this paper we will study if such a deviation can be accounted for in 2HDMs of types I, II, X, and Y. A contribution to $a_{\mu}$ able to bridge the $\Delta a_{\mu}$ discrepancy can arise in 2HDMs from a light pseudoscalar through Barr-Zee type two-loop diagrams~\cite{Chang:2000ii,Cheung:2001hz,Krawczyk:2002df,Larios:2001ma,Cheung:2003pw}. However, a light pseudoscalar may be in conflict with a heavy charged scalar whose mass is strongly constrained by direct and indirect searches. In fact, the general 2HDM lower bound on the mass of the charged scalar $H^{\pm}$ from direct searches at LEP2 is $M_{H^{\pm}} \gsim 79$~GeV~\cite{PDG2014}, and even stronger indirect bounds can be set from $B$-physics in type II and Y models.

In 2HDMs, the observed 126~GeV resonance can be identified with any of the two CP-even Higgs bosons.\footnote{In this paper, we work in the CP-conserving case \textit{i.e,} we assume all the parameters to be real. The CP-violating case (see \cite{Branco:2011iw} for a review) is interesting in its own right as it can significantly modify the phenomenology (see for example Ref.~\cite{Basso:2012st} and references therein). We will leave the CP-violating case for a future study.} 
In the present paper we identified this resonance with the lightest CP-even scalar $h$.  This interpretation is possible in all four 2HDMs types considered here. In particular, we chose the limit $\beta-\alpha=\pi/2$ in which the couplings of the light CP-even neutral Higgs $h$ with the gauge bosons and fermions attain the SM values.  In fact, the measured signal strengths and production cross section of such a particle are in very good agreement with the corresponding SM predictions~\cite{Chen:2013kt,Chen:2013rba,Baglio:2014nea, Eberhardt:2013uba,Baglio:2014nea,Cheng:2014ova,Barroso:2013zxa,Chang:2012ve,Belanger:2013xza,Barger:2013ofa,Chang:2013ona,Cheung:2013rva,Celis:2013ixa,Ferreira:2014sld}.

In addition to the bounds set by the muon $g$$-$$2$,  2HDMs are constrained by direct searches at colliders for the Higgs bosons $h, H, A$ and $H^\pm$, $B$-physics observables, EW precision measurements and theoretical considerations of vacuum stability and perturbativity. The question then arises: which of these models are preferred by the present set of direct and indirect constraints? In this work we addressed this question concentrating on the four models described in Table \ref{yukawas}. Our analysis shows that only models of Type X (``lepton specific") survive all these constraints.

The paper is organised as follows. In Sec.~II we present a detailed analysis of the EW constraints on the masses of the pseudoscalar boson $A$, charged scalar $H^\pm$, and additional neutral heavy scalar $H$. We study radiative corrections in the 2HDMs and, in particular, the impact of the precise measurements of the $W$ boson mass $M_W$ and the effective weak mixing angle $\sin^2\!\theta^{\text{lept}}_{\text{eff}}$. It is then important to check whether a large mass hierarchy between $A$ and $H^\pm$ is allowed by the Higgs measurements at the LHC and by the theoretical constraints on vacuum stability and perturbativity, which is discussed in Sec.~III. In Sec.~IV we present the additional contributions of the 2HDMs to the muon $g$$-$$2$ and discuss their implications on the four types of model analysed in this paper. Prospects for constraints from the electron $g$$-$$2$ are presented in Sec.~V. Conclusions are drawn in Sec.~VI.

\section{Electroweak constraints}
\label{sec:ewc}

In this section we analyze the constraints arising from EW precision observables on 2HDMs. In particular, we compare the theoretical 2HDMs predictions for $M_W$ and $\sin^2\!\theta^{\text{lept}}_{\text{eff}}$ with their present experimental values via a combined $\chi^2$ analysis~\cite{Broggio:MSthesis}.

As it was shown for the first time in~\cite{Sirlin:1980nh}, in the SM the $W$ mass can be computed perturbatively by means of the following relation
\begin{align}\label{mw}
	M^2_W = \frac{M^2_Z}{2}\left[1 + \sqrt{1 - \frac{4 \pi \alpha_{\rm em}}{\sqrt{2} G_F M^2_Z} \frac{1}{1 - \Delta r}} \, \right],
\end{align}
where $\alpha_{\rm em}$ is the fine-structure constant, $G_F$ is the Fermi constant and $M_Z$ is the $Z$ boson mass. The on-shell quantity $\Delta r$~\cite{Sirlin:1980nh}, representing the radiative corrections, is a function of the parameters of the SM. In particular, since $\Delta r$ also depends on $M_W$, Eq.~(\ref{mw}) can be solved in an iterative way. The relation between the effective weak mixing angle $\sin^2\!\theta^{\text{lept}}_{\text{eff}}$ and the on-shell weak mixing angle $\sin^2\!\theta_W$ is given by \cite{Gambino:1993dd}
\begin{align}
	\sin^2\!\theta^{\text{lept}}_{\text{eff}} = k_l \! \left(M^2_Z\right) \sin^2\!\theta_W \, ,
\end{align}
where $\sin^2\!\theta_W=1-M^2_W/M^2_Z$ \cite{Sirlin:1980nh} and $k_l(q^2) = 1+\Delta k_l(q^2)$ is the real part of the vertex form factor $Z\to l \bar{l}$ evaluated at $q^2 = M^2_Z$.

The 2HDM $\mathcal{O}(\alpha)$ corrections to $\Delta r$ and $\Delta k_{l}$ can be written in form
\begin{align}
\Delta r^{\mbox{$\scriptscriptstyle{\rm 2HDM}$}}\, =\,& \Delta r + \Delta \tilde{r}\, ,\\
\Delta k^{\mbox{$\scriptscriptstyle{\rm 2HDM}$}}_l \,=\,& \Delta k_l + \Delta \tilde{k}_l\, ,
\end{align}
where the tilded quantities indicate the additional 2HDM contributions not contained in the SM prediction. These additional corrections depend only on the particles and parameters of the extended Higgs sector which are not present in the SM part. The radiative corrections $\Delta r$ and $\Delta k_{l}$ are known up to two-loop order, including some partial higher-order EW and QCD corrections~\cite{Awramik:2003rn,Awramik:2006uz} (for a review of these corrections we refer the reader to~\cite{Sirlin:2012mh}). For our purposes, this level of accuracy in the SM part is not needed, and in our codes~\cite{Broggio:MSthesis} we implemented the full one-loop SM result plus the leading two-loop contributions of~\cite{Degrassi:1996ps,Degrassi:1997iy,Ferroglia:2002rg}. The additional correction $\Delta \tilde{r}$ has been available for a long time~\cite{Bertolini:1985ia}. We recalculated this contribution and found agreement with the previous results. The additional 2HDM correction $\Delta \tilde{k}_l$ was not available in the literature. We evaluated it following the notation of~\cite{Bertolini:1985ia}. For convenience, the calculation was carried out in the $\overline{\text{MS}}$ scheme and then translated to the on-shell scheme by means of the relations derived in~\cite{Degrassi:1990tu,Gambino:1993dd}. The analytic results can be found in~\cite{Broggio:MSthesis}. Following the analysis of~\cite{Bertolini:1985ia}, we neglected the $\mathcal{O}(\alpha)$ corrections where a virtual Higgs is attached to an external fermion line, since they are suppressed by factors of $\mathcal{O}(M_f/M_W)$. As a result, no new contributions to vertex and box diagrams are present with respect to the SM ones. All the additional diagrams fall in the class of bosonic self-energies and $\gamma$-$Z$ mixing terms. We point out that, in this approximation, these EW constraints do not depend on the way fermions couple to the Higgs bosons and, therefore, all four types of 2HDMs discussed in this paper share the same EW constraints.

The 2HDM predictions for $M_W$ and $\sin^2\!\theta^{\text{lept}}_{\text{eff}}$ depend on the $Z$ boson mass $M_Z=91.1876~(21)$ GeV~\cite{PDG2014}, the top quark mass, $m_t = 173.2~(0.9)$ GeV \cite{Aaltonen:2012ra}, the strong coupling constant $\alpha_s(M_Z)=0.1185~(6)$ \cite{PDG2014}, the variation of the fine-structure constant due to light quarks, $\Delta \alpha^{(5)}_{\text{had}}(M^2_Z) = 0.02763~(14)$~\cite{Hagiwara:2011af}, the masses of the neutral Higgs bosons $M_h=126$~GeV, $M_H$ and $M_A$, the charged Higgs mass $M_{H^\pm}$,  and the combination $(\beta -\alpha)$ of the mixing angles in the scalar sector, which we will set to $\pi/2$ to be consistent with the LHC results on Higgs boson searches~\cite{Chen:2013kt,Chen:2013rba,Baglio:2014nea, Eberhardt:2013uba,Baglio:2014nea,Cheng:2014ova,Barroso:2013zxa,Chang:2012ve,Belanger:2013xza,Barger:2013ofa,Chang:2013ona,Cheung:2013rva,Celis:2013ixa,Ferreira:2014sld}. To analyze the constraints on 2HDMs arising from the present measurements of $M_W$ and $\sin^2\theta^{\text{lept}}_{\text{eff}}$ we define
\be
	\chi_{\mbox{$\scriptscriptstyle{\rm EW}$}}^2=
	\left(\frac{M_W^{\mbox{$\scriptscriptstyle{\rm 2HDM}$}} - 
	M_W^{\mbox{$\scriptscriptstyle{\rm EXP}$}}}{\sigma_{M_W}^{\mbox{$\scriptscriptstyle{\rm EXP}$}}} \right)^2+
	\left(\frac{
	\sin^2\!\theta^{\text{lept}, \scriptscriptstyle{\rm 2HDM} }_{\text{eff}} - 
	\sin^2\!\theta^{\text{lept}, \scriptscriptstyle{\rm EXP} }_{\text{eff}}}
	{\sigma_{\sin^2\theta_W}^{\mbox{$\scriptscriptstyle{\rm EXP}$}}} \right)^2,
\ee
and use the following experimental values for $M_W$~\cite{PDG2014} and $\sin^2\!\theta^{\text{lept}}_{\text{eff}}$~\cite{ALEPH:2005ab}:
\beqa \label{exp}
	M_W^{\mbox{$\scriptscriptstyle{\rm EXP}$}} &=& 80.385 \pm 0.015~{\rm GeV}, \nonumber \\
	\sin^2\!\theta^{\text{lept}, \scriptscriptstyle{\rm EXP}}_{\text{eff}} &=& 0.23153 \pm 0.00016.
	\eeqa
We note that the corrections $\Delta r^{\mbox{$\scriptscriptstyle{\rm 2HDM}$}}$ and $\Delta k^{\mbox{$\scriptscriptstyle{\rm 2HDM}$}}_l$ implemented in our code receive a large contribution from the well-known quantity $\Delta \rho^{\mbox{$\scriptscriptstyle{\rm 2HDM}$}}=\Delta \rho+\Delta \tilde{\rho}$:
\begin{align}\label{drho1}
	\Delta r^{\mbox{$\scriptscriptstyle{\rm 2HDM}$}} \, =& ~~\Delta \alpha^{\mbox{$\scriptscriptstyle{\rm 2HDM}$}}
	- \frac{\cos^2\!\theta_W}{\sin^2\!\theta_W} \Delta \rho^{\mbox{$\scriptscriptstyle{\rm 2HDM}$}} + \ldots \, ,\\
\label{drho2}
	\Delta k_l^{\mbox{$\scriptscriptstyle{\rm 2HDM}$}} \, =& 
	+\frac{\cos^2\!\theta_W}{\sin^2\!\theta_W} \Delta \rho^{\mbox{$\scriptscriptstyle{\rm 2HDM}$}} + \ldots \, ,
\end{align}
 where $~\Delta \alpha^{\mbox{$\scriptscriptstyle{\rm 2HDM}$}}$ is the photon vacuum polarization contribution in the 2HDM. The definition of the parameter $\Delta\rho$, consistent with Eqs.~(\ref{drho1},\ref{drho2}), can be found in \cite{Jegerlehner:1991ed}.

\begin{figure}[!ht]
\centering
\includegraphics[width=1.0\textwidth]{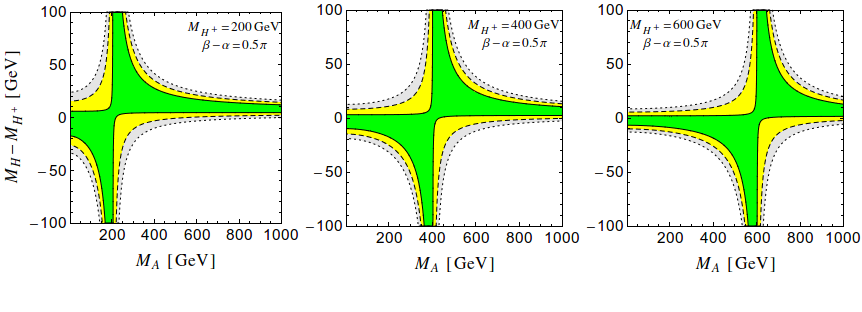}
\caption{The parameter space allowed in the $M_A$ vs.\ $\Delta M_H =M_H-M_{H^\pm}$ plane by EW precision constraints ($M_W$ and $\sin^2\theta^{\text{lept}}_{\text{eff}}$). The green, yellow, gray regions satisfy $\Delta \chi_{\mbox{$\scriptscriptstyle{\rm EW}$}}^2 (M_A, \Delta M)< 2.3, 6.2, 11.8$, corresponding to 68.3, 95.4, and 99.7\% confidence intervals, respectively. From left to right, different values of $M_{H^\pm}= 200,~400$ and 600 GeV are shown. All plots employ $\beta-\alpha = \pi/2$ and $M_h = 126$~GeV, 
and $M_Z$, $m_t $, $\alpha_s(M_Z)$ and $\Delta \alpha^{(5)}_{\text{had}}(M^2_Z)$ are set to their measured central values.}
\label{fig:ewpc}
\end{figure}

The results of our analysis are displayed in Fig.~\ref{fig:ewpc}, where we chose three different values of the charged scalar mass, $M_{H^\pm} = 200$, 400 and $ 600$ GeV, the Higgs-to-gauge boson coupling $\beta-\alpha = \pi/2$, $M_h=126$~GeV, and we set 
$M_Z$, $m_t $, $\alpha_s(M_Z)$ and $\Delta \alpha^{(5)}_{\text{had}}(M^2_Z)$ to their experimental central values. The green, yellow and gray regions of the plane $M_A$ vs.\ $\Delta M_H =M_H-M_{H^\pm}$ where drawn requiring
$
	\Delta \chi_{\mbox{$\scriptscriptstyle{\rm EW}$}}^2 (M_A, \Delta M) = 
	\chi_{\mbox{$\scriptscriptstyle{\rm EW}$}}^2 (M_A, \Delta M) - \chi_{\mbox{$\scriptscriptstyle{\rm EW}$},{\rm min}}^2 
	< 2.3, 6.2, 11.8,
$
respectively, which are the critical values corresponding to the 68.3, 95.4, and 99.7\% confidence intervals ($\chi_{\mbox{$\scriptscriptstyle{\rm EW}$},{\rm min}}^2$ is the absolute minimum of $\chi_{\mbox{$\scriptscriptstyle{\rm EW}$}}^2 (M_A, \Delta M)$)~\cite{PDG2014, Rolke:2004mj}. Note that in the case of a large splitting between $M_H$ and $M_{H^\pm}$, $M_A$ is required to be almost degenerate with $M_{H^\pm}$ in order to satisfy the EW constraints. This point has already been remarked upon in~\cite{Eberhardt:2013uba,Celis:2013ixa}. In addition, we observe that all values of $M_A$ are allowed when $M_H$ and $M_{H^\pm}$ are almost degenerate. This useful result will be used in Sec.~\ref{sec:gm2}.
%
%

\section{Theoretical Constraints on the splitting $M_A$-$M_{H^+}$}
\label{sec:thc}

Although, as shown in the previous section, any value of $M_A$ is allowed by the EW precision tests in the limit of $M_H \sim M_{H^\pm}$, a large separation between $M_{H^\pm}$ and $M_A$ is strongly constrained by theoretical considerations of vacuum stability and perturbativity. Since we are interested in a light pseudoscalar (motivated by the resolution of  the muon $g$$-$$2$ discrepancy), it is important to check how small $M_A$ is allowed to be. In this section we study such constraints in a semi-analytical way.

The CP-conserving 2HDM with softly broken $Z_2$ symmetry is parametrized by seven real parameters, namely $\lambda_{1,..,5}$, $m_{12}^2$ and $\tan\beta$~\cite{Gunion:2002zf,Branco:2011iw}. The general scalar potential of two Higgs doublets $\Phi_1$ and $\Phi_2$ is given by 
\begin{eqnarray} \label{scalar-potential}
	V &=& m_{11}^2 |\Phi_1|^2 
	+ m_{22}^2 |\Phi_2|^2 - m_{12}^2 (\Phi_1^\dagger \Phi_2 + \Phi_1 \Phi_2^\dagger) \nonumber \\
	&& + {\lambda_1\over2} |\Phi_1|^4 + {\lambda_2 \over 2} |\Phi_2|^4 
	+ \lambda_3 |\Phi_1|^2 |\Phi_2|^2 + \lambda_4 |\Phi_1^\dagger \Phi_2|^2 + {\lambda_5 \over 2} 
	\left[ (\Phi_1^\dagger \Phi_2)^2 + (\Phi_1 \Phi_2^\dagger)^2\right],
\end{eqnarray}
where the Higgs vacuum expectation values are given by  $ \langle \Phi_{1,2} \rangle = \frac{1}{\sqrt{2}}(0, v_{1,2})^T$. The masses of all the physical Higgs bosons and the mixing angle $\alpha$ between CP-even neutral ones are obtained from $\tan\beta$ and the remaining six real parameters~\cite{Gunion:2002zf}. The vacuum stability and perturbativity conditions put bounds on these parameters and correlate the masses of different neutral and charged scalars. For example, the vacuum stability condition requires~\cite{Gunion:2002zf}:
\be \label{vacuum-stability}
	\lambda_{1,2}>0,~~\lambda_3>-\sqrt{\lambda_1
	\lambda_2},~~|\lambda_5|<\lambda_3+\lambda_4+\sqrt{\lambda_1 \lambda_2}, 
\ee
and the requirement of global minimum is imposed by the condition~\cite{Barroso:2013awa}
\be \label{global-minimum}
 	m_{12}^2(m_{11}^2-m_{22}^2\sqrt{\lambda_1/\lambda_2})(\tan\beta-(\lambda_1/\lambda_2)^{1/4})>0~,
 \ee
where $m_{11}$ and $m_{22}$ are functions of $\lambda_i$, $m_{12}$ and $\tan\beta$ as expressed in Ref.~\cite{Gunion:2002zf}. 
For the perturbativity criterion, we will consider three different values for the maximum couplings
\be \label{perturbativity}
	|\lambda_i| \lesssim |\lambda_{\rm max}| =  \sqrt{4\pi}, 2\pi, 4\pi,
\ee
to see their impact on the allowed mass spectrum. A large separation between any two scalar masses in 2HDM is controlled by the above constraints.

\begin{figure}[!ht]
\centering
\subfigure{\includegraphics[width=0.4\textwidth]{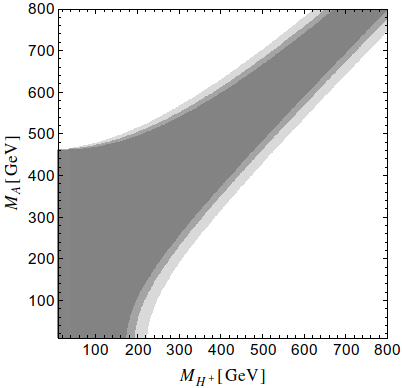}}\quad
\subfigure{\includegraphics[width=0.4\textwidth]{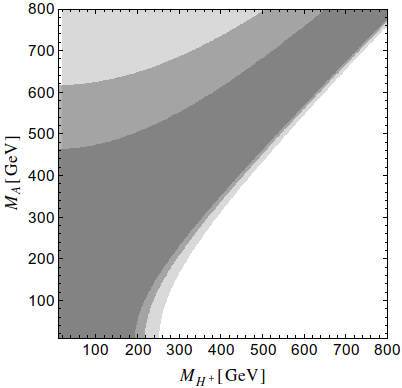}}
\caption{Theoretical constraints on the $M_A$-$M_{H^\pm}$ plane in 2HDMs with softly broken $Z_2$ symmetry. We employ $\beta-\alpha = \pi/2$ and $M_h=126$ GeV. The darker to lighter gray regions in the left panel correspond to the allowed regions for $\Delta M\equiv M_H-M_{H^\pm}= \{20,0,-30\}$ GeV and $\lambda_{\rm max} = \sqrt{4 \pi}$. The allowed regions in the right panel correspond to $\lambda_{\rm max} = \{\sqrt{4 \pi}, 2 \pi, 4 \pi\}$ and vanishing $\Delta M$. Both plots are obtained for $\tan\beta = 50$, but the change with respect to values of $\tan\beta \in [5,100]$ is negligible.}
\label{fig:semiana}
\end{figure}

For a given value of $\tan\beta$, one can express two of the six parameters, namely $\lambda_4$ and $\lambda_5$, entirely in terms of physical masses $M_A$, $M_{H^\pm}$ and the parameter $m_{12}$ using the relations \cite{Gunion:2002zf}
\beqa \label{mA-mHp}
	M_A^2 &=& \frac{m_{12}^2}{\sin\beta \cos\beta}-\lambda_5 \upsilon^2, \nonumber \\
	M_{H^\pm}^2 &=& M_A^2+\frac{1}{2}\upsilon^2(\lambda_5 - \lambda_4). 
\eeqa
Furthermore, for a given value of $\tan\beta$ and solving for the $M_h$, $M_H$ and the SM-like Higgs coupling limit $\beta-\alpha=\pi/2$, one can obtain semi-analytical solutions for the remaining four real parameters in terms of four physical masses and the only free parameter $\lambda_1$ using the expressions given in Ref. \cite{Gunion:2002zf}. The expressions for $\lambda_{2,3}$ valid for $\tan\beta \gg 1$ are
\beqa \label{eq:tb10}  \label{eq:tb55}
	\lambda_2 \upsilon^2 & \simeq &  M_h^2 + \lambda _1 \upsilon^2/\tan^4\beta, \nonumber \\
	\lambda_3 \upsilon^2 & \simeq &  2 M_{H^\pm}^2-  2 M_{H}^2 + M_h^2 + \lambda _1
	\upsilon^2/\tan^2\beta, \nonumber \\
	m_{12}^2 &\simeq & M_H^2/\tan\beta + ( M_h^2 -\lambda_1 v^2)/\tan^3\beta \,.
\eeqa
We find that in the $\beta-\alpha=\pi/2$ limit the parameters $\lambda_{2,3}$ depend negligibly on $\tan\beta$. Similar expressions for $\lambda_{4,5}$ can be obtained using Eq.~(\ref{mA-mHp}). One can now impose the conditions (\ref{vacuum-stability}), (\ref{global-minimum}) and (\ref{perturbativity}) on the above equations. As can be seen from Eq.~(\ref{mA-mHp}), the difference 
$M_{H^\pm}^2 -M_A^2 $ is proportional to   $\lambda_5-\lambda_4$ and it is restricted to be smaller than  $\lambda_3+\sqrt{\lambda_1 \lambda_2}$ as required by vacuum stability condition, Eq.~(\ref{vacuum-stability}). Both $\lambda_2$ and $\lambda_3$ have almost negligible dependence on $\lambda_1$ as can be seen from the semi-analytic expressions in Eqs.~(\ref{eq:tb10}). Taking $M_h=126$ GeV, $\lambda_1=\lambda_{\rm max}$ and imposing all the theoretical constraints mentioned above, one gets the regions allowed in $M_A$-$M_{ ^\pm}$ plane as shown in Fig.~\ref{fig:semiana}.
%

The plots in Fig.~\ref{fig:semiana} depend very mildly on $\tan\beta$ so that similar results hold for any value of $\tan\beta \in [5,100]$. We also note that the change in the allowed regions is negligible with respect to small departures from the SM-like Higgs coupling limit $\beta-\alpha = \pi/2$. One can clearly see that for a light pseudoscalar with $M_A \lesssim 100$~GeV the charged Higgs boson mass gets an upper bound of $M_{H^\pm} \lesssim 200$~GeV. Also, Fig.~\ref{fig:semiana} shows the presence of lower bounds on $M_A$ if the charged Higgs boson mass is heavier than $\sim 200$~GeV. We will discuss the implications of these correlations in the following sections.

\section{Constraints from the muon $g-2$}
\label{sec:gm2}

The SM prediction of the muon $g$$-$$2$ is conveniently split into QED, EW and hadronic contributions:
$
    a_{\mu}^{\mysmall \rm SM} = 
         a_{\mu}^{\mysmall \rm QED} +
         a_{\mu}^{\mysmall \rm EW}  +
         a_{\mu}^{\mbox{$\scriptscriptstyle{\rm H}$}}.
$  
The QED prediction, computed up to five loops, currently stands at
$a_{\mu}^{\mysmall \rm QED} = 116584718.951\,(80)
\times 10^{-11}$~\cite{Aoyama:2012wk}, 
while the EW effects provide
$a_{\mu}^{\mysmall \rm EW} = 153.6\,(1.0) \times 10^{-11}$~\cite{Czarnecki:1995wq, Czarnecki:1995sz, Gnendiger:2013pva}.
The latest calculations of the hadronic leading order contribution, via the hadronic $e^+ e^-$ annihilation data,
are in agreement:
$a_{\mu}^{\mbox{$\scriptscriptstyle{\rm HLO}$}}  =$ 
$6903\,(53) \times 10^{-11}$~\cite{Jegerlehner:2009ry},
$6923\,(42) \times 10^{-11}$~\cite{Davier:2010nc} and 
$6949\,(43) \times 10^{-11}$~\cite{Hagiwara:2011af}.
The next-to-leading order hadronic term is further divided into two
parts:
$
     a_{\mu}^{\mbox{$\scriptscriptstyle{\rm HNLO}$}}=
     a_{\mu}^{\mbox{$\scriptscriptstyle{\rm HNLO}$}}(\mbox{vp})+
     a_{\mu}^{\mbox{$\scriptscriptstyle{\rm HNLO}$}}(\mbox{lbl}).
$
The first one, 
$-98.4\, (7) \times 10^{-11}$~\cite{Hagiwara:2011af},
is the $O(\alpha^3)$ contribution of diagrams containing hadronic vacuum polarization insertions~\cite{Krause:1996rf}. The
second term, also of $O(\alpha^3)$, is the leading hadronic light-by-light contribution; the latest calculations of this term,
$105\,(26) \times 10^{-11}$\cite{Prades:2009tw}
and
$116\,(39) \times 10^{-11}$\cite{Jegerlehner:2009ry},
are in good agreement, and an intense research program is under way to improve its evaluation~\cite{Colangelo:2014dfa,Colangelo:2014pva,Blum:2014oka,Pauk:2014rfa}. Very recently, also the next-to-next-to leading order hadronic corrections have been determined: insertions of hadronic vacuum polarizations were computed to be 
$a_{\mu}^{\mbox{$\scriptscriptstyle{\rm HNNLO}$}}(\mbox{vp}) = 12.4\,(1) \times 10^{-11}$~\cite{Kurz:2014wya},
while hadronic light-by-light corrections have been estimated to be
$a_{\mu}^{\mbox{$\scriptscriptstyle{\rm HNNLO}$}}(\mbox{lbl}) = 3\,(2) \times 10^{-11}$~\cite{Colangelo:2014qya}.
If we add the value $a_{\mu}^{\mbox{$\scriptscriptstyle{\rm HLO}$}}  = 6903\,(53) \times 10^{-11}$ of \cite{Davier:2010nc} (which roughly coincides with the average of the three hadronic leading order values reported above) to the conservative estimate $a_{\mu}^{\mbox{$\scriptscriptstyle{\rm HNLO}$}}(\mbox{lbl})=116\,(39) \times 10^{-11}$ of~\cite{Jegerlehner:2009ry} and the rest of the other {\small SM} contributions, we obtain
\be
  a_{\mu}^{\mbox{$\scriptscriptstyle{\rm SM}$}}= 116591829 \, (57) \times 10^{-11}
\ee
(for reviews of $a_{\mu}^{\mbox{$\scriptscriptstyle{\rm SM}$}}$ 
see~\cite{Jegerlehner:2009ry,Blum:2013xva,Melnikov:2006sr,Davier:2004gb,Passera:2004bj,Knecht:2003kc}). The difference between $a_{\mu}^{\mbox{$\scriptscriptstyle{\rm SM}$}}$ and the experimental value~\cite{Bennett:2006fi}
\be
    a_{\mu}^{\mbox{$\scriptscriptstyle{\rm EXP}$}}  = 116592091 \, (63) \times 10^{-11}
\ee
is, therefore,
$\Delta a_{\mu} \equiv a_{\mu}^{\mbox{$\scriptscriptstyle{\rm EXP}$}}-a_{\mu}^{\mbox{$\scriptscriptstyle{\rm SM}$}} = +262 \, (85) \times 10^{-11}$,
i.e.\ 3.1$\sigma$ (all errors were added in quadrature).

Models with two Higgs doublets give additional contributions to $a_\mu$ which could bridge the above discrepancy $\Delta a_{\mu}$~\cite{Chang:2000ii,Cheung:2001hz,Larios:2001ma,Krawczyk:2002df,Cheung:2003pw}. All the Higgs bosons of the 2HDMs contribute to $a_{\mu}$. However, in order to explain $\Delta a_{\mu}$, their total contribution should be positive and, as we will see, enhanced by $\tan \beta$. In the 2HDM, the one-loop contributions to $a_{\mu}$ of the neutral and charged Higgs bosons are~\cite{Lautrup:1971jf, Leveille:1977rc,Dedes:2001nx} 
\be
	\delta a_\mu^{\mbox{$\scriptscriptstyle{\rm 2HDM}$}}({\rm 1loop}) = 
	\frac{G_F \, m_{\mu}^2}{4 \pi^2 \sqrt{2}} \, \sum_j  \left (y_{\mu}^j \right)^2  r_{\mu}^j \, f_j(r_{\mu}^j),
\label{amuoneloop}
\end{equation} 
where $j =  \{h, H, A , H^\pm\}$, $r_{\mu}^ j =  m_\mu^2/M_j^2$, and 
\begin{eqnarray}
	f_{h,H}(r) &=& \int_0^1 \! dx \,  { x^2 ( 2- x) \over 1 - x + r x^2},  
\label{oneloopintegrals1} \\
	f_A (r) &=& \int_0^1 \! dx \,  { -x^3  \over 1 - x +r  x^2},  
\label{oneloopintegrals2} \\
	f_{H^\pm} (r) &=& \int_0^1 \! dx \, {-x (1-x)  \over 1 - (1-x) r}.
\label{oneloopintegrals3}
\end{eqnarray} 
The normalized Yukawa couplings $y_{\mu}^{h, H,A}$ are listed in Table \ref{yukawas}, and $y_{\mu}^{H^{\pm}} \!\! = y_{\mu}^A$. The one-loop contribution of the light CP-even boson $h$ is given by~\eq{amuoneloop} with $j=h$; however, as we work in the limit $\beta - \alpha \approx \pi/2$ in which $h$ has the same couplings as the SM Higgs boson, its contribution is already contained in $a_{\mu}^{\mysmall \rm EW}$ and shouldn't therefore be included in the additional 2HDM contribution (in any case, this contribution is negligible: setting $M_h=126$~GeV and $y_{\mu}^h =1$ we obtain  $2 \times 10^{-14}$). The formulae in Eqs.~(\ref{amuoneloop}--\ref{oneloopintegrals3}) show that the one-loop contributions to $a_{\mu}$ are positive for the neutral scalars $h$ and $H$, and negative for the pseudo-scalar and charged Higgs bosons $A$ and $H^{\pm}$ (for $M_{H^\pm} > m_{\mu}$). In the limit $r\ll1$, 
\begin{eqnarray}
	f_{h,H}(r) &=&- \ln r - 7/6 + O(r), 
	\label{oneloopintegralsapprox1} \\
	f_A (r) &=& +\ln r +11/6 + O(r),
	\label{oneloopintegralsapprox2} \\
	f_{H^\pm} (r) &=& -1/6 + O(r),
	\label{oneloopintegralsapprox3}
\end{eqnarray} 
showing that in this limit $f_{H^\pm}(r)$ is suppressed with respect to $f_{{h,H,A}}(r)$.

The one-loop results in Eqs.(\ref{oneloopintegralsapprox1}--\ref{oneloopintegralsapprox3}) also show that, in the limit $r\ll1$, $\delta a_\mu^{\mbox{$\scriptscriptstyle{\rm 2HDM}$}}({\rm 1loop})$ roughly scales with the fourth power of the muon mass. For this reason, two-loop effects may become relevant if one can avoid the suppression induced by these large powers of the muon mass. This is indeed the case for two-loop Barr-Zee type diagrams with effective $h\gamma \gamma$, $H\gamma \gamma$ or  $A\gamma \gamma$ vertices generated by the exchange of heavy fermions~\cite{Chang:2000ii}. Their contribution to the muon $g$$-$$2$ is~\cite{Czarnecki:1995wq,Chang:2000ii,Cheung:2001hz,Cheung:2003pw}
\be
	\delta a_\mu^{\mbox{$\scriptscriptstyle{\rm 2HDM}$}}({\rm 2loop-BZ}) = \frac{G_F \, m_{\mu}^2}{4 \pi^2 \sqrt{2}} \, \frac{\alpha_{\rm em}}{\pi} 
	\, \sum_{i,f}  N^c_f  \, Q_f^2  \,  y_{\mu}^i  \, y_{f}^i \,  r_{f}^i \,  g_i(r_{f}^i),
\label{barr-zee}
\end{equation} 
where $i = \{h, H, A\}$, $r_{f}^i =  m_f^2/M_i^2$, and $m_f$, $Q_f$ and $N^c_f$ are the mass, electric charge and number of color degrees of freedom of the fermion $f$ in the loop. The functions $g_i(r)$ are
\be
\label{2loop-integrals}
	g_i(r) = \int_0^1 \! dx \, \frac{{\cal N}_i(x)}{x(1-x)-r} \ln \frac{x(1-x)}{r},
\ee
where ${\cal N}_{h,H}(x)= 2x (1-x)-1$ and ${\cal N}_{A}(x)=1$.
As in the one-loop case, the two-loop Barr-Zee contribution of the light scalar $h$ is given by the formula in~\eq{barr-zee} with $i=h$ but, once again, working in the limit $\beta - \alpha \approx \pi/2$, its contribution is already contained in $a_{\mu}^{\mysmall \rm EW}$ and we will therefore not include it in the additional 2HDM contribution (setting $M_h=126$~GeV, $y_{\mu}^h = y_f^h=1$ and summing over top, bottom and tau lepton loops we obtain $-1.4 \times 10^{-11}$). Note the enhancement factor $m_f^2/m_{\mu}^2$ of the two-loop formula in~\eq{barr-zee} relative to the one-loop contribution in~\eq{amuoneloop}. As this factor $m_f^2/m_{\mu}^2$ can overcome the additional loop suppression factor $\alpha / \pi$,  the two-loop contributions in~\eq{barr-zee} may  become larger than the one-loop ones. Moreover, the signs of the two-loop functions $g_{h,H}$ (negative) and $g_{A}$ (positive) for the CP-even and CP-odd contributions are opposite to those of the functions $f_{h,H}$ (positive) and $f_{A}$ (negative) at one-loop. In type II models in the limit $\beta-\alpha=\pi/2$, a numerical calculation shows that for a light scalar with mass lower than $\sim 5$~GeV and $\tan \beta \gsim 10$ the negative two-loop scalar contribution is larger than the positive one-loop result; also, for $M_A \gsim 3$~GeV  and $\tan \beta \gsim 5$ the positive two-loop pseudoscalar contribution is larger than the negative one-loop result. A light pseudoscalar with $M_A \gsim 3$~GeV can therefore generate a sizeable positive contribution which can account for  the observed $\Delta a_{\mu}$ discrepancy.\footnote{One could also advocate a very light scalar with mass lower than $\sim 5$~GeV, but this scenario is challenged experimentally~\cite{Franzini:1987pv} (see also \cite{Schmidt-Hoberg:2013hba}).} A similar conclusion is valid for the pseudoscalar contribution in type X models \cite{Cao:2009as}. In fact, we notice from the pseudoscalar Yukawa couplings in Table \ref{yukawas} that the contribution of the tau lepton loop is enhanced by a factor $\tan^2 \! \beta$ both in type II and in X models; on the contrary, it is suppressed by $1/\tan^2 \! \beta$ in models of type I and Y.

\begin{figure}[!ht]
\centering
\subfigure{\includegraphics[width=0.4\textwidth]{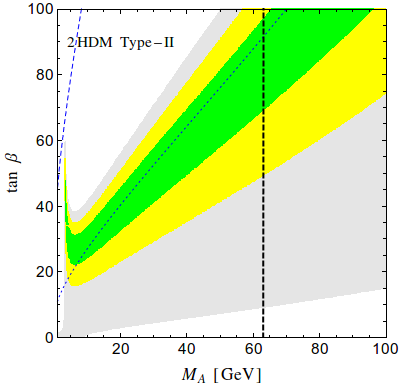}}\quad
\subfigure{\includegraphics[width=0.4\textwidth]{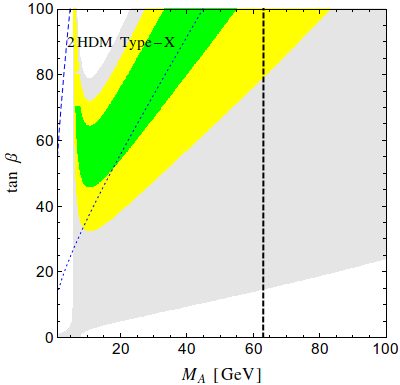}}
\caption{The $1\sigma$, $2\sigma$ and $3\sigma$ regions allowed by $\Delta a_\mu$ in the
$M_A$-$\tan\beta$ plane taking the limit of $\beta-\alpha=\pi/2$ and $M_{h(H)}=126$ (200) GeV in
type II (left panel) and type X (right panel) 2HDMs. The regions below the dashed (dotted) lines are allowed at 3$\sigma$ (1.4$\sigma$) by $\Delta a_e$. The vertical
dashed line corresponds to $M_A ~= ~m_h/2$ (see text for an explanation). }
\label{fig:g-2}
\end{figure}

The additional 2HDM contribution $\delta a_{\mu}^{\mbox{$\scriptscriptstyle{\rm 2HDM}$}}  = \delta a_\mu^{\mbox{$\scriptscriptstyle{\rm 2HDM}$}}({\rm 1loop}) + \delta a_\mu^{\mbox{$\scriptscriptstyle{\rm 2HDM}$}}({\rm 2loop-BZ})$ obtained adding Eqs.~(\ref{amuoneloop}) and~(\ref{barr-zee}) (without the $h$ contributions) is compared with $\Delta a_\mu$ in Fig.~\ref{fig:g-2} for type II and X models as a function of tan$\beta$ and $M_A$. Once again, we used the SM coupling limit $\beta-\alpha=\pi/2$. In both models, relatively small $M_A$ values are needed to generate the positive pseudoscalar contribution to $a_{\mu}$ required to bridge the $\Delta a_{\mu}$ discrepancy. In turn, in order to satisfy the theoretical constraints of Sec.~\ref{sec:thc} for a light pseudoscalar with $M_A \lsim 100$~GeV, the charged Higgs mass must be lower than $\sim 200$~GeV, as shown in Fig.~\ref{fig:semiana}, but anyway larger that the model-independent LEP bound $M_{H^{\pm}} \gsim 79$~GeV~\cite{PDG2014}. Under these conditions, the EW constraints discussed in Sec.~\ref{sec:ewc} restrict the value of the neutral scalar mass to be $M_H \sim M_{H^\pm}$ (see Fig.~\ref{fig:ewpc}). We therefore chose the conservative values $M_H = M_{H^\pm} =200$~GeV to draw Fig.~\ref{fig:g-2}. Slightly higher values of $\tan \! \beta$ would be preferred in Fig.~\ref{fig:g-2} if the lower values $M_H = M_{H^\pm} =150$~GeV were chosen instead (in fact, a lower $M_H$ induces a slightly larger negative scalar contribution to $a_\mu$). For given values of $M_A$ and $\tan\beta$, the contribution to $\delta a_{\mu}^{\mbox{$\scriptscriptstyle{\rm 2HDM}$}}$  in type II models is slightly higher than that in type X models because of the additional $\tan^2 \! \beta$ enhancement for the down-type quark contribution. It is important to note that, on the contrary, type I and Y models cannot account for the present value of $\Delta a_\mu$ due to their lack of $\tan^2 \! \beta$ enhancements.
%
%

In 2HDMs of type II (and Y) a very stringent limit can be set on $M_{H^\pm}$ from the flavour observables Br$({\bar B} \to X_s \gamma)$ and $\Delta m_{B_s}$, as well as from the hadronic $Z \to b\bar{b}$ branching ratio $R_b$: $M_{H^\pm} > 380$ GeV at 95\% CL irrespective of the value of $\tan \! \beta$~\cite{Misiak:2006zs, Deschamps:2009rh, Mahmoudi:2009zx, Hermann:2012fc}. This bound is much stronger than the model-independent one obtained at LEP, $M_{H^{\pm}} > 79.3$~GeV at 95\% CL~\cite{PDG2014,Heister:2002ev}. This strong constraint $M_{H^\pm} > 380$~GeV, combined with the theoretical requirements shown in Fig.~\ref{fig:semiana}, leads to $M_A \gtrsim 300$~GeV. In turn, this lower bound on $M_A$ is in conflict with the required value for $\Delta a_\mu$, as can be seen from  Fig.~\ref{fig:g-2}. Therefore, type II models are strongly disfavoured by these combined constraints. On the other hand, no such strong flavour bounds on $M_{H^\pm}$ exist in type X models~\cite{Branco:2011iw,Mahmoudi:2009zx}. These models are therefore consistent with all the constraints we considered, provided $M_A$ is small and $\tan\! \beta$ large (see Fig.~\ref{fig:g-2}), $M_{H^{\pm}}  \lsim 200$~GeV (Fig.~\ref{fig:semiana}), and $M_H \sim M_{H^\pm}$ (Fig.~\ref{fig:ewpc}).

Finally, it has been pointed out in Ref. \cite{Cao:2009as} that the decay $h \to A A$ could have a large branching fraction in the limit cos($\beta-\alpha$) $\approx ~0$ and tan$\beta \gg~$1,
for sufficiently light pseudo scalar masses. While we have not performed a detailed analysis of this process, we would like to point out that any limits from this decay can be
avoided by considering the region $m_A ~> ~m_h/2 $. From Fig. (\ref{fig:g-2}), we see that there is still sufficient amount of the parameter space left which can provide an
explanation to the excess in the $(g-2)_\mu$.  

\section{Constraints from the electron $g-2$}
\label{sec:electron}

It is usually believed that new-physics contributions to the electron $g$$-$$2$ ($a_e$) are too small to be relevant; with this assumption, the measurement of $a_e$ is equated with the SM prediction to determine the value of the fine-structure constant $\alpha_{\rm em}$: 
$a_e^{\mbox{$\scriptscriptstyle{\rm SM}$}}(\alpha_{\rm em}) = a_e^{\mbox{$\scriptscriptstyle{\rm EXP}$}}$.
However,  as discussed in~\cite{Giudice:2012ms}, in the last few years the situation has been changing thanks to several theoretical~\cite{Aoyama:2012wj} and experimental~\cite{Hanneke:2008tm} advancements in the determination of $a_e$ and, at the same time, to new independent measurements of  $\alpha_{\rm em}$ obtained from atomic physics experiments~\cite{Bouchendira:2010es}. The error induced in the theoretical prediction $a_e^{\mbox{$\scriptscriptstyle{\rm SM}$}}(\alpha_{\rm em})$ by the experimental uncertainty of $\alpha_{\rm em}$ (used as an input, rather than an output), although still dominating, has been significantly reduced, and one can start to view $a_e$ as a probe of physics beyond the SM.

The present difference between the SM prediction  and the experimental value is
$\Delta a_e \equiv a_e^{\mbox{$\scriptscriptstyle{\rm EXP}$}}-a_e^{\mbox{$\scriptscriptstyle{\rm SM}$}} = -10.5 (8.1) \times 10^{-13}$,
i.e.\ 1.3 standard deviations, thus providing a beautiful test of QED. We note that the sign of $\Delta a_e$ is opposite to that of $\Delta a_{\mu}$ (although the uncertainty is still large). The uncertainty $8.1\times 10^{-13}$ is dominated by that of the SM prediction through the error caused by the uncertainty of $\alpha_{\rm em}$, but work is in progress to reduce it significantly~\cite{Terranova:2013vfa}. Following the analysis presented above for the muon $g$$-$$2$, we compared $\Delta a_e$ with the 2HDM contribution $\delta a_e^{\mbox{$\scriptscriptstyle{\rm 2HDM}$}}  = \delta a_e^{\mbox{$\scriptscriptstyle{\rm 2HDM}$}}({\rm 1loop}) + \delta a_e^{\mbox{$\scriptscriptstyle{\rm 2HDM}$}} ({\rm 2loop-BZ})$ obtained adding Eqs.~(\ref{amuoneloop}) and~(\ref{barr-zee}), obviously replacing $m_{\mu}$ and $y_{\mu}^j$ with $m_e$ and $y_e^j$. The result is shown once again in Fig.~\ref{fig:g-2}, for type II and X models, as a function of tan$\beta$ and $M_A$. In each panel, the region below the dashed (dotted) line is the $3\sigma$ ($1.4\sigma$) region allowed by $\Delta a_e$. Clearly, the precision of $\Delta a_e$ is not yet sufficient to play a significant role in limiting the 2HDMs, but this will change with new, more precise, measurements of $a_e$ and $\alpha_{\rm em}$. For example, reducing the uncertainty of $\Delta a_e$ by a factor of three and maintaining its present (negative) central value, the $3\sigma$ regions allowed by $\Delta a_e$ completely disappear from both panels of Fig.~\ref{fig:g-2}. In fact, at present, increasing by $1\sigma$ the negative central value $\Delta a_e = -10.5 \times 10^{-13}$, one still gets a negative value, which cannot be accounted for in the region of parameter space shown in Fig.~\ref{fig:g-2}. (Increasing the present central value by $1.4 \sigma$ one gets $+0.8 \times 10^{-13}$, which is the input used to draw the dotted lines in Fig.~\ref{fig:g-2}.) Obviously, future tests will depend both on the uncertainty and on the central value of $\Delta a_e$.

\section{Conclusions}
  
In recent times there has been renewed interest in the phenomenology of models with two Higgs doublets. Most of the focus
 has been on four possible variations of them, namely, type I, II, X (or ``lepton specific") and Y (``flipped"). In this work we presented a detailed phenomenological analysis with the aim of challenging these four models. We included constraints from electroweak precision tests, vacuum stability and perturbativity, direct searches at colliders, muon and electron $g$$-$$2$, and constraints from $B$-physics observables. In these models, all the Higgses couple similarly to the gauge bosons, but differently to the fermions.  Therefore, the electroweak constraints (along with the perturbativity and vacuum stability ones) are common to all of them, while the rest of the constraints vary from model to model. Using a stringent set of precision electroweak measurements we showed that, in the limit $(\beta - \alpha) \to {\pi/2}$ consistent with the LHC results on Higgs boson searches, all values of $M_A$ are allowed when $M_H$ and $M_{H^\pm}$ are almost degenerate. We considered a CP-conserving scenario where the 126~GeV resonance discovered at the LHC has been identified with the lightest CP-even boson $h$.

The 2HDM predictions for observables which depend on fermion couplings are expected to vary from model to model. In fact, the interplay between the muon $g$$-$$2$ and $b \to s \gamma$ is the key distinguisher between the various types. A light pseudoscalar with couplings proportional to $\tan \! \beta$ is required to explain the discrepancy between the SM prediction and the observed value of the muon $g$$-$$2$. This is only possible in type II and X models. On the other hand, in type II and Y models the BR($b \to s \gamma$) sets a strong $\mathcal{O}$(380~GeV) lower bound on the mass of  the charged scalar which, taken together with the perturbativity and vacuum stability constraints, was shown to leave hardly any space for a light pseudoscalar. On the contrary, only loose constraints arise from the BR($b \to s \gamma$) in type I and X models, because both up and down type quarks couple to the same Higgs doublet in these models. Therefore, we showed that type X (``lepton specific") models are the only ones which can accommodate the muon $g$$-$$2$ without violating the BR$(b \to s + \gamma)$ and the rest of the present constraints. We also noted that an improved measurement of the electron $g$$-$$2$ may lead in the future to further significant bounds on 2HDMs.

The parameter space favourable for the muon $g$$-$$2$ in type X models is quite limited in mass ranges for the heavy neutral and charged scalar: $M_H \sim M_{H^\pm} \lesssim 200$ GeV (with small $M_A$ and large $\tan\! \beta$). These bosons can be searched for in forthcoming collider experiments, even if this parameter region could be elusive because the productions of the additional Higgs bosons $A, H$, and $H^\pm$ are suppressed either by $1/\tan^2\beta$ (in single productions, e.g.\ through gluon fusion) or by $\cos(\beta-\alpha)$ (associated productions of $V\phi$ and $h\phi$). The leading search channels for the extra bosons would then be pair or associated productions through $pp \to \gamma/Z/W \to H^+ H^-/HA/H^\pm A/H^\pm H$ followed by the decays $H^\pm \to l^\pm \nu$ and $A,H\to l^+ l^-$, which can be readily tested at the next run of the LHC \cite{Branco:2011iw,Kanemura:2014bqa}.

~

\textbf{Acknowledgements}
We would like to thank G.~Degrassi, T.~Dorigo, A.~Ferroglia, P.~Paradisi, A.~Sirlin and G.~Venanzoni for very useful discussions.
MP and KMP also thank the Department of Physics and Astronomy of the University of Padova for its support. Their work was supported in part by the PRIN 2010-11 of the Italian MIUR and by the European Program INVISIBLES (PITN-GA-2011-289442).
MP and SKV thank the hospitality of KIAS, Seoul, where this work  started.  SKV also thanks DST, Govt.\ of India, for the support through the project SR/S2/RJN-25/2008.

\bibliography{2hdm}

\end{document}